# Photoinduced dynamics in protonated aromatic amino acid

G. Grégoire[3], B. Lucas[1], M. Barat[1], J. A. Fayeton[1], C. Dedonder-Lardeux[2] and C. Jouvet[*2]
[1]*Laboratoire des Collisions Atomiques et Moléculaires, (CNRS UMR C8625), Bât. 351, Univ. Paris-Sud 91405 Orsay Cedex, France*
[2]*Laboratoire de Photophysique Moléculaire (CNRS UPR 3361) Bât. 210, Univ. Paris-Sud 91405 Orsay Cedex, France*
[3]*Laboratoire de Physique des Lasers (CNRS UMR 7538), Institut Galilée, Université Paris 13, 93430 Villetaneuse, France*

## Abstract

UV photoinduced fragmentation of protonated aromatics amino acids have emerged the last few years, coming from a situation where nothing was known to what we think a good understanding of the optical properties. We will mainly focus this review on the tryptophan case. Three groups have mostly done spectroscopic studies and one has mainly been involved in dynamics studies of the excited states in the femtosecond/picosecond range and also in the fragmentation kinetics from nanosecond to millisecond. All these data, along with high level *ab initio* calculations, have shed light on the role of the different electronic states of the protonated molecules upon the fragmentation mechanisms.

## I. Introduction

Among naturally occurring amino acids, three possess fluorescent aromatic chromophores: tryptophan (Trp), tyrosine (Tyr) and phenylalanine (Phe). Trp has been the most studied owing to its strong fluorescence yield and the rich variety of information it provides according to the possible environments[1]. It is used as a natural fluorescent tag of proteins and was therefore investigated to obtain structural information of tryptophanyl proteins. Nevertheless, the knowledge of the primary photoinduced processes occurring in the bare molecule is nevertheless required to understand the difference observed in the fluorescence studies.

UV photo-excitation studies of gas phase protonated amino acids molecules have emerged in recent years with the combination of mass spectrometry and laser spectroscopy techniques. Several experimental groups have investigated in both frequency(ref 1,2,3) and time(ref 7,9) domains the fragmentation of these biomolecules after UV absorption. For protonated aromatic amino acids, UV photo-induced dissociation (PID) mass spectrometry reveals some of the non-radiative deactivation channels that cannot be directly addressed by means of fluorescence spectroscopy. Among the non-radiative pathways, inter- system crossing (ISC), chemical reaction, isomerisation, internal conversion (IC) and bond cleavage, the two latter processes, namely IC and H-atom loss, have been evidenced and distinguished in the case of protonated tryptamine[2]. Through IC, the UV photon energy (4.66 eV at 266 nm) is converted into internal energy within the electronic ground state. This internal energy is large enough to induce fragmentations. For instance, less than 2 eV only are required for an ammonia loss[3]. This excess energy is randomized and consequently leads to a statistical-type dissociation of the ion on the ground state potential energy surface, as also observed in collision-induced dissociation (CID) experiments. On the other hand, during the H-atom loss reaction, most of the energy initially put in the excited state is imparted to the leaving hydrogen atom after the amino N-H bond cleavage. The elimination of the hydrogen atom is a direct and fast process and results in the formation of a radical cation that exhibits a specific fragmentation pattern.

Three experimental approaches have been recently developed. Nanosecond UV spectroscopy on protonated molecules [4-6] in which the fragmentation yield as a function of the excitation wavelength has been investigated. Coincidence experiments allow to identify the neutral fragment(s) associated with a particular ionic channel [7-9] which imposes severe constraints on the possible fragmentation mechanisms. Moreover, it has been possible to extract the fragmentation times which are less than 10 ns for all primary fragmentation channels and in the μs range for some secondary fragment. Femtosecond pump/probe experiments [2,10,11] have revealed the excited state dynamics. Moreover it has been shown that the probe photon can change the branching ratio between the competitive fragmentation channels.

These are flexible molecules that have a number of stable ground state conformers even at low temperature. The lowest-energy protonation site for aromatic amino acids is on the amino group, while protonation on the indolic nitrogen for tryptophan lies about 1 eV higher in energy. $TrpH^+$ exhibits low-lying rotamers (figure 1) that mainly differ in the orientation of the amino-acid moiety above the indole chromophore by $2\pi/3$ rotation along of the $C_\alpha$-$C_\beta$ bond and by $\pi$ rotation along the $C_\gamma$-$C_\beta$ bond where $C_\gamma$ is the carbon atom of the indole ring. Protonated tyrosine and phenylalanine also present the same kind of stable rotamers and these different structures have to be included in the excited state calculations.

Excited states *ab-initio* calculations[6,10,12,13] have shown the very important role of the low lying excited states in protonated aromatic amino acids. Beside the two well-known electronic states of indole, $L_a$ and $L_b$ of $\pi\pi^*$ nature lie low lying charge transfer states, in which the excited electron is localized on the amino acid part either in a $\sigma^*$ orbital on the amino group or in a $\pi^*$ orbital localized on the CO group (here the symmetry plane is the glycine plane). P. R. Callis *et al.* have shown that the fluorescence of protein containing tryptophyl residue can be understood by the quenching of the $\pi\pi^*$ state due to charge transfer to the C=O peptide backbone[14-16]. A simple model [10] has been able to predict the role of the $\pi\sigma^*$ state to control the excited lifetime for different protonated amino acids: the energy of this $\pi\sigma^*$ state is dependent on the IP of the neutral molecule. Vertical and adiabatic excited state calculations [12] at the CC2 level have confirmed this crude model and have explored several deactivation pathways in the excited state: H loss, H transfer to carbonyl, proton transfer to the indole ring.

The present review is devoted to show how information coming from four experimental groups and two theoretical groups can be brought together to disentangle the fragmentation dynamics combining spectroscopic and dynamics studies of fragmentation in a large time scale, ranging from hundreds of femtoseconds to millisecond. From such information the role of the different electronic states and that of the conformation on the fragmentation pattern can now be understood.

## II. Experimental observations

II.1. Spectroscopic studies

The UV spectroscopy of protonated tryptophan has been investigated by three teams. At room temperature, the spectrum starts at 290 nm (4.27 eV) and the presence of another electronic state at 265 nm (4.65 eV) is evidenced. The fragmentation pattern changes with the energy and in particular the m/z= 130 fragment ($C_\alpha$-$C_\beta$ bond rupture) ion signal increases with the excess energy [5]. At lower temperature (liquid nitrogen), the 0-0 transition appears more clearly at 284.5 nm (4.35 eV) but is still broad [6]. At even lower temperature (5K) the spectrum is still broad whereas in similar conditions the protonated tyrosine spectrum exhibits narrow vibronic features[4]. This clearly emphasizes the presence of a low lying predissociative excited state in protonated tryptophan.

The observation of the m/z=204 channel (H loss) (see figure 2) and *ab-initio* calculations have been the key to understand the possible deactivation pathways[6,12,13,17]. At the same energy as the locally excited state where the excited electron is in a π* orbital on the indole ring, there is a πσ* state where the electron is localized on the protonated amino group. The presence of the excess electron on this group leads to the formation of a hypervalent species unstable along the NH bond. The presence of this state is also responsible for the very short excited state lifetime measured for protonated tryptamine[2]. The variation of the energy gap between the ππ* state (on the aromatic moiety) and the πσ* state with the aromatic chromophore enables to rationalise very simply the difference of the excited lifetime observed between tryptophan (hundred of fs at 4.6 eV) and tyrosine (tens of ps at 4.6 eV)[10]. In a few words, the energy of this πσ* state is the energy needed to remove an electron from the aromatic ring minus the electronic affinity of the protonated amino group (ionization potential of the $NH_4$ radical). The IP of indole being lower than the phenol IP, the πσ* state is lower in energy in tryptophan than in tyrosine and the coupling of the ππ* to the dissociative πσ* state is more efficient.

If this simple picture is basically correct it does not give a full comprehensive mechanism of the dynamics of the protonated aromatic amino acid, which is in fact rather complex and very rich as further experiments and *ab initio* calculations have shown.

II.2. Coincidences experiment

The development of a new experiment in which both the ionic and the neutral fragments issued from the photo-excitation of protonated aromatic amino acid are detected in coincidence has shed a new light on the fragmentation mechanism and brought much new information on the dynamics of these systems. The experiment by itself and the interpretation of the data is quite complex and can be found in very recent publications[8,9,18]. We will just recall here the main results deduced from these experiments: Measurement of the fragmentation time constants for each fragmentation channel, detection of the neutral(s) partner(s) associated with a fragment ion and deduction of the sequence of the fragmentation process in the case of many body dissociation scheme.

1) Except for the m/z=130 fragment issued from the H loss channel, all the fragmentation processes are finished in 10 μs after excitation with the 266 nm photon. The primary fragmentation events leading to the formation of m/z=188, 159, 132 take place within 10 ns[1]. Secondary fragmentations (leading to m/z=146/144) occur in less than 10 μs.

[1] A first series of experiments has been performed in a box held at a fixed voltage and indicate that all primary fragmentation events occur within 150 ns. These experiments have been repeated in an electric field as in the Lucas et al experiment[14] and the reaction time deduced is of the order of 10 ns.

2) The fragmentation of the tryptophan radical cation (m/z=204) obtained after the H loss spans a much broader time scale, from μs up to the ms range, and leads to the $C_\alpha$-$C_\beta$ bond rupture (m/z=130).

3) The fragmentation parent/daughter relationship can be tracked without ambiguity with the coincidence experiment [9]

a) m/z=205 → m/z=188 +$NH_3$ (in less than 10 ns) → m/z=146 +$CH_2CO$ (μs)
→ m/z =144+ $CO_2$ (μs)
b) m/z=205 → m/z=187 + $H_2O$ → m/z=159 + CO (in less than 10 ns)
c) m/z=205 → m/z=132 + NH2CH2COOH (less than 10 ns)[8]
d) m/z=205 → m/z=204 + H → m/z=130 + NH2CH2COOH (from μs to ms)
e) m/z=205 → m/z=130 + NH2CH2COHOH (less than 10ns, excited state)[7]

A few comments on these results are necessary channel a). This process occurs in the ground state after a complex rearrangement [19] and has a barrier of 1 eV[3]. This is the lowest energy channel. The energy content of the remaining ion after $NH_3$ loss is such that the fragmentation can go on and end up to the formation of m/z=144 or 146 on a longer time scale.

Channel b). This process probably occurs in the ground state after a proton transfer from the amino group toward the acidic group, as depicted in the case of collision induced dissociation (CID) experiments (ref siu et ohair). The first step is the water loss and the CO loss occurs immediately after since the CO is very weakly bond after the water loss, with a barrier to dissociation less than 0.1eV.

Channel c). This fragmentation channel was quite unexpected. From CID experiments, one would expect this fragment to be issued from the HCN loss (m/z= 27) from the m/z=159. The coincidence experiment demonstrates however that this fragment is produced in one step only. A mechanism has been proposed to rationalize this observation: proton transfer from the $NH_3$ group toward the indolic cycle, followed by the $C_\alpha$-$C_\beta$ bond rupture and formation of a the ionic complex in which there is an hydrogen transfer from the amino group of the glycine to the nitrogen of the pyrrolic group [8].

Channel e) recent experiments have shown [7] that the $C_\alpha$-$C_\beta$ bond rupture, which has been assigned to the fragmentation of the m/z=204 radical cation after the H loss, is also produced through a direct fragmentation in the excited state. The coincidence experiments have shown that the fragmentation leading to m/z=130 occurs on two time scales: a fast process (less than 10 ns) which is observed in tyrosine and tryptophan but not in tyramine and tryptamine and a slow process (a few μs) observed for the all four species. Through ab initio calculations, a direct "barrier free mechanism" has been found for the fast process: the first step is the electron transfer from the indole ring to the amino acid part followed by a proton transfer from the amino group toward the carbonyl group and then a direct (barrier free) dissociation along the $C_\alpha$-$C_\beta$ bond in the excited state [7].

## II.3. Femtoscecond pump/probe experiment

The pump(266nm)/probe(800nm) signal obtained for protonated tryptophan, which measures the excited state lifetime, is quite complex *since the time evolution recorded* is strongly dependent on the fragment channel as shown on figure 3. As it will be seen the full interpretation of such a complex signal requires much more information than just the lifetime measurement[11]. In particular it should require what are the possible fragmentation mechanisms and the parent/daughter ion links that have been obtained very recently with the experiment presented above[8,9].

We here recall the main results:

1) The number of fragment ions is conserved in the pump/probe scheme. The energy in the system after the UV pump excitation is such that all the excited ions break within the time window of the experiment (about 10 μs), and the IR probe laser does not induce new fragmentation pathways. The probe photon changes the branching ratio between the different fragmentation channels.

2) The pump/probe scheme measures the excited state lifetime of the protonated species before any internal conversion process. However, when a radical cation is formed after hydrogen loss (m/z=204), then its open shell structure allows further absorption of the IR probe laser that enhances its specific secondary fragment at mass m/z= 130, as seen for protonated tryptamine[2] .

3) It is possible to fit all the data with a combination of two exponential decays of 400 fs and 15 ps. However having two time constants does not necessarily mean that there are two species having two well defined lifetimes [20]. Probably the distribution of conformers is such that the excited population has an uniform (broad) distribution of lifetimes between (few fs from the spectroscopic measurement[4]) to a couple of tens ps and such a distribution can be fitted by a bi-exponential decay as shown in the Gregoire et al paper[20].
It should be mentioned that the broadening observed in the excitation spectra, which predicts a lifetime shorter than 10 fs, and the pump/probe experiment, which exhibit longer lifetimes (hundreds of fs) are not in contradiction since the very short lived species cannot be re-excited by the probe photon and thus cannot be detected.

4) Depending on the recorded fragmentation channel, the effect of the probe laser upon the fragmentation efficiency differs:

a) for the m/z=132 channel no pump/probe signal is observed

b) all the final products (m/z=146,144,159) are depopulated upon the 800 nm photon absorption especially on the 15 ps time constant part.

c) the m/z=130 is populated by the 800 nm absorption at short time, medium, long time. The long time behaviour has been well characterised in the case of tryptamine[2] as the absorption of the radical cation issued from the hydrogen loss.

d) the m/z=188 fragment, which is the primary fragment of ions at masses m/z=146 and m/z=144, is depopulated at short time (400 fs), populated at medium time (15 ps). This unexpected behaviour is the most difficult to understand and imposes severe constraints on the possible fragmentation mechanism.

## III. Excited states *ab initio* calculations

### III.1. Protonated tryptophan

The *ab initio* calculations and particularly the excited state optimizations at the couple clusters level (CC2) have been of a great help to understand the deactication pathways in protonated tryptophan. From these calculations some general features emerge:

Three excited states with different electronic configurations are involved in the protonated species. Depending on the molecules and on their conformers, the ordering of these excited states differs. The $\pi\pi^*$ state where the excited electron is localized on the indole ring bears the oscillator strength for UV excitation. In the same energy range lies a $\pi\sigma^*$ resulting from a transfer of the electron from the $\pi^*$ indole ring orbital (indole being the plane of symmetry) toward a $\sigma^*$ orbital on the amino part of the amino acid. This $\sigma_{NH3}^*$ orbital has a $\sigma$ symmetry with respect to the amino acid plane. The third kind of state lies higher in energy and can be characterized as a charge transfer state where the electron is localised on a $\pi^*_{CO}$ orbital with respect to the amino acid plane. Note that within the TD-DFT framework, such charge transfer state is predicted to be the lowest singlet state, which turns to be an artefact of the DFT method[17,21].

We have already emphasized[12] that the exact nature of the electronic states is difficult to assert due to the strong mixing of the orbitals. Besides, geometry optimizations of the different excited states have revealed that the deactivation pathways are strongly dependent upon the nature of the electronic state, *i.e.* wherever the electron is localized. When the system evolves through a $\pi\pi^*$ state, then a direct proton transfer from the protonated amino group towards the aromatic chromophore occurs. Such mechanism leads to internal conversion process with fragmentation channel occurring in the electronic ground state. Whereas when the electron is transferred to the amino acid chain, then the excess electron triggers specific reactions in the excited state such as H atom loss (from a $\pi\sigma^*$ state) and H atom transfer to the carbonyl (from the $\pi^*_{CO}$ state).

In order to get a clearer interpretation of the influence of these charge transfer states upon the excited state dynamics and the role of the probe photon upon the deactivation pathways, we present below the case of hydrogenated glycine, which can be viewed as the simplest system able to mimic such electron transfer in the protonated aromatic amino acids.

III.2. Hydrogenated glycine

The optimization of the protonated glycine has been obtained at the ri-mp2 level with the cc-pVDZ basis set. We have calculated the energy of the excited states of the hydrogenated species starting from the geometry of the protonated form in Cs geometry (figure 5).

The two first electronic states $D_0$ and $D_1$ can be viewed as the charge transfer states calculated in the protonated aromatic amino acids that can be exited through UV pump photon. In the ground state the unpaired electron of the hydrogenated glycine is localized on the $\sigma$ orbital on the $NH_3$. The optimization of the hydrogenated molecule on this ground state structure leads to the lengthening of the NH bond, which could lead eventually to the H loss (as shown for Tryptamine) if the symmetry is broken (D0). In the first a'' excited state (D1), the unpaired electron is delocalized on the $\pi^*$ orbital on the carbon chain and on the oxygen of the carbonyl. Optimization of this state leads to a barrier free H transfer towards the C=O. This process is thus thought to be very fast.

The $D_2$ and $D_3$ excited states lie higher in energy (between 1.5 - 1.9 eV) and can be viewed as the final states populated through 800nm probe photon in the aromatic amino acids. Therefore, one has to consider them in order to rationalize the effect of the probe laser upon the fragmentation channels. The optimization of the second a" state (D2) in which the electron is localized on the 6a" orbital does not lead to large geometry changes (in the Cs geometry) as compared to the protonated optimized structure. However this orbital is clearly antibonding along the two NH out of plane coordinates. One expects that it should lead to H loss even more efficiently than the ground state. The optimization of the first a' excited state (D3) (18a' orbital which is antibonding along the C…$NH_3$ bond) leads to an increase of the $C_\alpha$…NH3 bond length and is probably dissociative along this coordinate (at longer CN distance, the SCF convergence cannot be achieved).

It should be stressed that internal conversion can occur along the dissociation paths in the excited state and brings the system back to the ground state with a lot of internal energy. The fragmentation channels will thus be similar to the ones observed in CID[3,19,22]. However, one might expect that the branching ratio between the different fragmentation channels changes drastically. In low-energy collision experiments, the excess energy is sequentially brought to the molecular system. The mobile proton model has been introduced to rationalize the different fragmentation channels observed with CID experiments. The proton moves along the peptide chain, weakens the peptide bond that leads to a/b/y-type of fragments. In contrary, the internal conversion process releases the total photon energy in one step and fragmentation can thus occur prior full energy redistribution over the degrees of freedom of the molecule (IVR).

## IV. Interpretation of the fs pump/probe signal

The meaning of the pump/probe signal has been extensively discussed in the case of tryptamine[2]. Briefly, if no probe photon is absorbed, the initial population evolving in an electronic surface will decay with a lifetime $\tau$ and the fragments are produced with given rate constants $k_i$ and a branching ratio $\alpha_i = \frac{k_i}{\sum_i k_i}$ (1). When a 800 nm photon is absorbed the energy in the system has increased by the probe photon energy and the fragments are produced with rate constants $k'_i$ and a branching ratio $\beta_i = \frac{k_i^{'}}{\sum_i k_i^{'}}$ (2). The population re-excited by the probe laser has been removed from the initially excited state and thus the pump/probe signal recorded for a fragment *i* is given by the difference of reactivity of the two pathways, with and without the probe laser:

$$S_i(t) = (\beta_i - \alpha_i)\exp(-t/\tau) \quad (3)$$

Equation (3) clearly points out that:

i) If the rate constants do not change with the internal energy of the hot molecule ($k_i = k'_i$), absolutely no time dependent signal is observed.

ii) If one channel is clearly dominant $((k_i, k^{'}_i) >> (k_j, k^{'}_j))$ then $\alpha_i \approx \beta_i \approx 1$ and also no dynamics is observed.

iii) A population initially created through UV pump excitation and evolving in a specific electronic state with a time constant $\tau$ can be promoted through IR probe absorption into a higher excited state with a different electronic character and then a different reaction path. This will lead to a depletion of the fragmentation yield of this initial fragmentation channel and a corresponding increase of the fragment ion intensity into another one.

In the case of protonated tryptophan, there are several low lying excited states with different electronic characters that can be excited with the 266 nm pump photon. As stated above, each of this electronic state is expected to decay with a specific time constant $\tau$ and can lead to different fragmentation channels. In order to interpret the pump/probe signal, we thus have to understand the effect of the probe photon when the system evolves on different electronic states and its influence on the reaction paths.

(i) For some conformers, the dynamical process will be: $\pi\pi^* \rightarrow \pi\pi^*_{CO} \rightarrow$ H transfer towards the carbonyl. This barrier free reaction is associated with the short time constant at 400 fs. The concerted electron-proton transfer to the carbonyl weakens the $C_a$-$C_b$ bond and will lead to the fragmentation channel at mass m/z=130.

If Internal Conversion occurs before fragmentation in the excited state, then the molecule becomes protonated on the carboxylic group. Such structure is thought to give rise to the formation of the immonium ion as demonstrated by Ohair[19]. If the proton moves back to the amino group before fragmentation, then ammonia loss happens along with its secondary fragments at masses m/z= 146 and 144.

The probe photon can be absorbed before the H transfer towards D2 (5a"-6a") transition (the oscillator strength is 10 times larger than the 5a"-18a' transition). From there one expects a fast H loss from the $NH_3$ group due to the anti bonding character of the orbital. The effect of the probe laser upon the fragmentation yield of the different channel is thus an increase of the signal at mass m/z= 130 and a depletion of all others fragmentation channels that can be produced after UV excitation through this deactivation path, *i.e.* at masses m/z= 188, 146, 144 and 159.

(ii) For other conformers, the process will be: $\pi\pi^* \rightarrow \pi\sigma^*$. Since there is a barrier to the H loss reaction, the lifetime of the $\pi\sigma^*$ state is expected to be longer (15 ps) than the $\pi_{CO}^*$ state. This deactivation pathways will lead to the formation of the radical cation at mass m/z= 204 along with its secondary fragment at mass m/z= 130.

If internal conversion happens along the H loss reaction, then the proton comes back to the amino group. The fragmentation channels are the ones introduced above, *i.e.* at masses m/z= 188, 146, 144 and 159.

The probe photon can be absorbed before the H loss reaction in the excited state. From the $\sigma^*$ orbital (17a' orbital in hydrogenated glycine (D0)), the probe photon can reach two excited states, D2 and D3. D2, as explained previously, leads to the H loss. Therefore, the reaction path is the same except that the total excess energy is increased by the energy of the probe photon. Therefore, the effect of the probe laser will be an increase of the signal at mass m/z=130.

Whereas D3 leads to the $NH_3$ loss and will lead to an increase of the signal at mass m/z =188 as experimentally observed. It should be mentioned that this fragmentation is quite different from the one issued from an internal conversion process. In this direct process occurring in the excited state, the energy content in the fragment is much less (excitation energy minus the CN bond dissociation around 3.5eV) than in the ground state process and most of the energy should be dissipated as kinetic energy of the lightest $NH_3$ fragment. This direct dissociation channel in the excited state is not expected to lead to secondary fragmentation channel at masses m/z=146 and 144.

For these two reaction pathways, one electron of the $\pi$ cloud of the indole chromophore has been transferred into the amino acid chain. The indole chromophore has thus an open shell electronic structure which allows 800 nm probe photon absorption. This will lead to the warming up of the system but not changing the pathway on the excited state dissociation.

(iii) The conformers for which the coupling towards the charge transfer states is weak will evolve on the ππ* electronic manifold. From ab initio calculations, this channel leads to a barrier free proton transfer towards the indole ring. This reaction path is thought to be very fast and will lead directly to the m/z=132 fragment in the ground state. (ref Jacqueline)

This is the only channel insensitive to the probe laser with a pump/probe signal completely flat. This imposes severe constraints on the dynamics in the excited state and in the kinetics of fragmentation for this channel. Before the proton transfer reaction to the indole chromophore has occurred, the indole ππ* electronic configuration should allow optical transition through higher ππ* states and charge transfer states. In that latter case, a decrease of this proton transfer reaction yield is expected but is not experimentally seen. Therefore, one has to conclude that this proton transfer reaction is faster than the temporal laser width (100 fs). In contrary, when the proton transfer reaction is completed, the only optical transition allowed with the 800 nm probe photon corresponds to ππ* excitation that would only heat up the system but not change the fragmentation channel.

Since the fragmentation leading to the mass m/z=132 happens on the ground electronic surface, one might expect that the other fragmentation channels enter the competition, in particular in the mobile proton model framework.[3] The absence of pump/probe signal on this channel clearly indicates that this is not the case and that the kinetics of fragmentation is faster than the proton exchange and total energy redistribution. It should be stressed that the coincidence experiment gives an upper limit for the kinetics of dissociation at less than 10 ns.

The overall picture of the excited state dissociation pathways is presented in figure 6 and summed up in the followings:

(i) The $\pi\pi^*_{CO}$ state leads to a barrier free H transfer from the $NH_3^+$ group towards the carbonyl group within 400 fs. The concerted electron-proton transfer to the carbonyl weakens the $C_\alpha$-$C_\beta$ bond and will lead to the fragmentation channel at mass m/z=130.

(ii) The πσ* state leads to the H loss through a small barrier within 15 ps. This leads to the formation of the radical cation at mass m/z=204 that can further fragment into its secondary fragment ion at mass m/z=130.

(iii) The ππ* state leads to a barrier free proton transfer from the $NH_3^+$ group towards the indole ring. The fragment ion associated with this proton transfer reaction is m/z=132.

## V. Conclusions.

The photofragmentation of protonated aromatic amino acids is a very rich and complex process. If some fragmentation channels are produced after internal conversion to the ground state surface, others are issued from direct dissociation in the excited states in a non-statistical process. It seems that there is a close relationship between the nature of the excited state (and the localization of the electron) and the fragmentation pathways. Since the ordering of the excited state is dependent on the conformation of the amino acid, the non-statistical fragmentation pathways will also depend on the initial geometry.

It seems also that, in the case of protonated tryptophan, some initial conformations lead to one specific channel (channel m/z=132) through internal conversion following proton transfer from the amino group towards the indole ring. This channel is representing an exception to the commonly assumed mobile proton model in which the proton transfer is faster than the fragmentation itself.

The coupling between the local excitation on the indole ring and the excited state in which the electron is localized on the amino acid part is the key to the fragmentation.

This implies that from this localized electron it is possible with a probe photon to reach low dissociative states and to drive specific fragmentations with a combination of optical pulses.

On larger system, the absence of coupling between the aromatic ring and the protonated amino acid part leads to a strong decrease of the fragmentation efficiency[23,24]. At the opposite in dipeptides [25,26] and tripeptides[27] a good coupling leads to a very fast decay of the excited state, similar to the one observed in tryptophan and tyrosine.

At last it seems that due to this relation it might be possible in the future to predict the conformation of a given peptide by controlling its fragmentation with optical pulses.

Figures captions:

Figure 1: lowest energy conformer of protonated tryptophan. The rotamers corresponds to $2\pi/3$ rotation along the $C_\alpha$-$C_\beta$ bond and $\pi$ rotation along the $C_\alpha$-$C_\beta$-$C_\gamma$ chain.

Figure 2: 266nm photofragment mass spectrum of protonated tryptophan at m/z= 205. Among the fragmentation channel, ion at mass m/z= 204 reveals the H loss reaction specifc to UV photoexcitation.

Figure 3: Time evolution of the pump/probe signal recorded at different m/z. The transients can be fitted by a combination of a short exponential decay (400fs) and a longer one (15ps). On m/z=132 no pump/probe signal is observed.

Figure 4: Orbitals involved in the first excited states of protonated tryptophan (drawn using Molekel software)

Figure 5: Excited states energy (eV) of hydrogenated radical glycine with the relevant molecular orbital (calculated with Turbomole at ri-cc2 level and cc-pVDZ basis set). The vertical arrows represent the electronic transition (and their transition moments in length jauge) from the first two lower electronic states.
Bottow: Highest doubly occupied orbital of the protonated glycine.
Right side: orbitals populated by the unpaired electron of the different electronic states.
Left side: most probable dissociation channel of the different electronic states.

Figure 6
Scheme of the fragmentation mechanisms for the various dissociation pathways, depending on the localization of the electron after laser excitation. Adapted from Lepere et al[9]

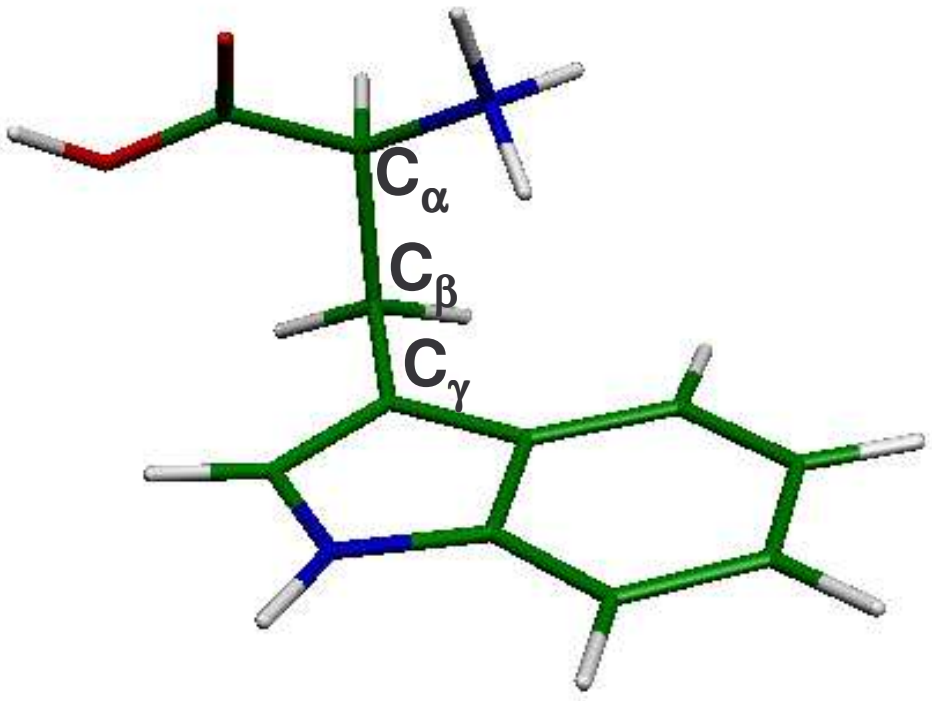

Figure 1

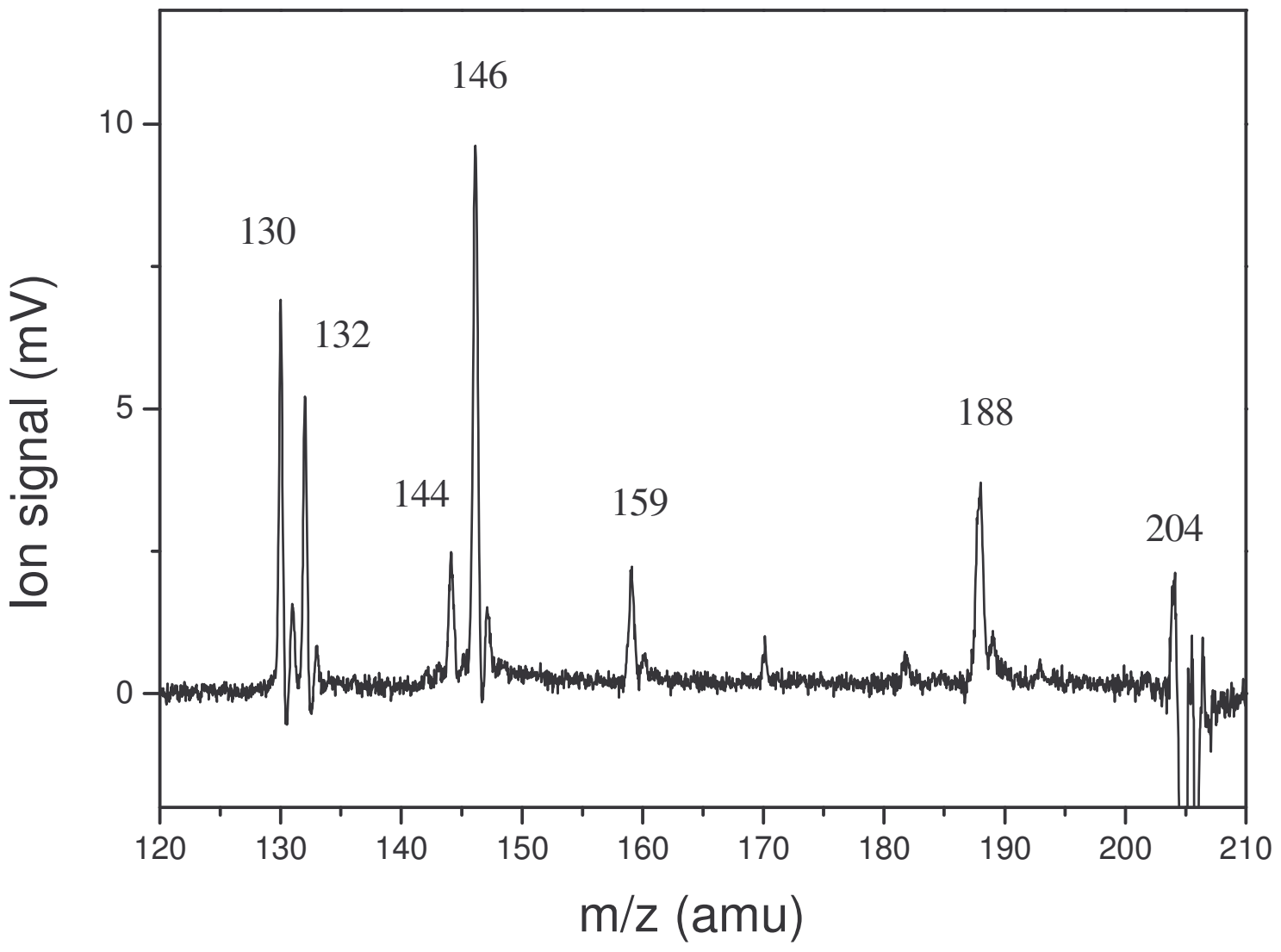

Figure 2

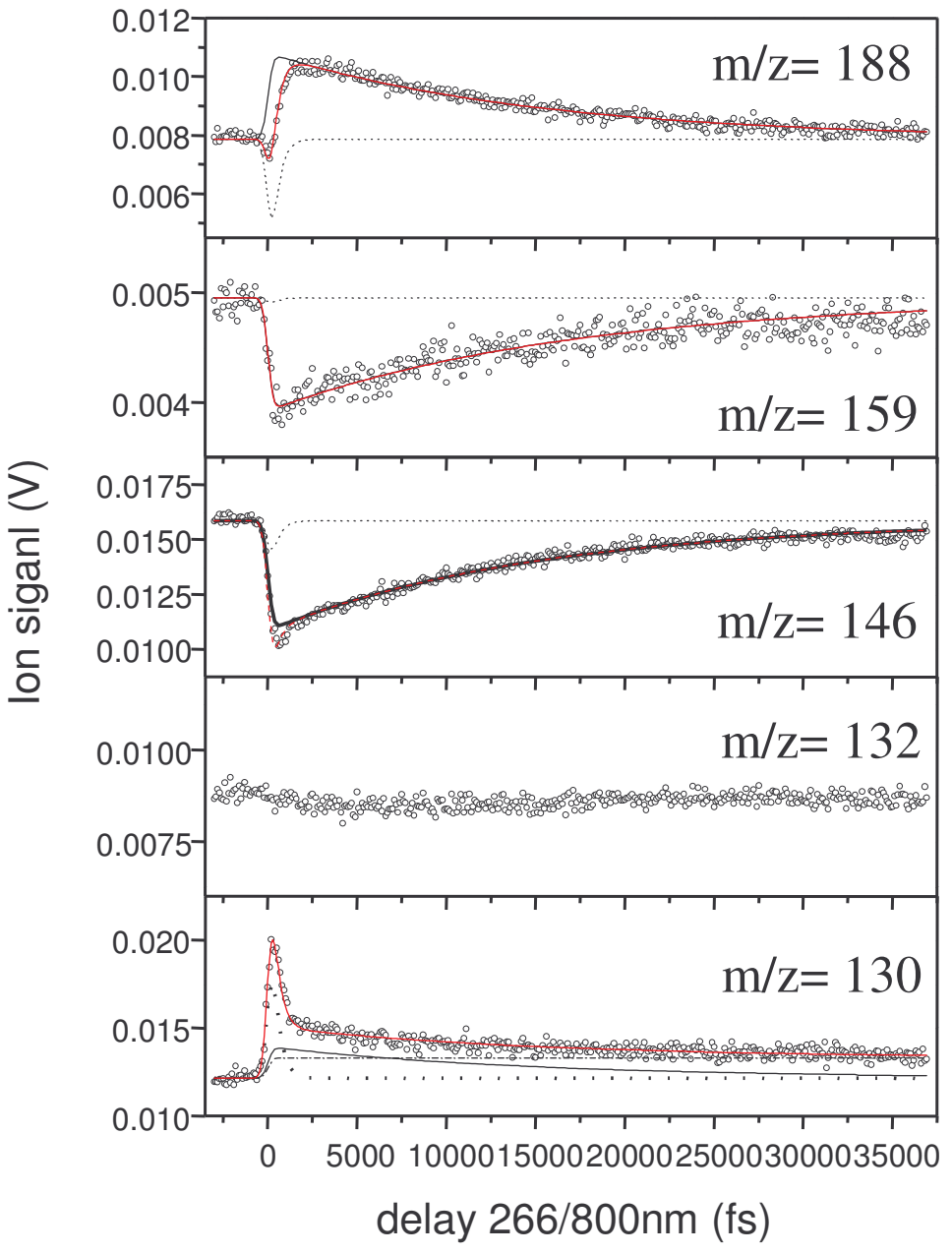

Figure 3

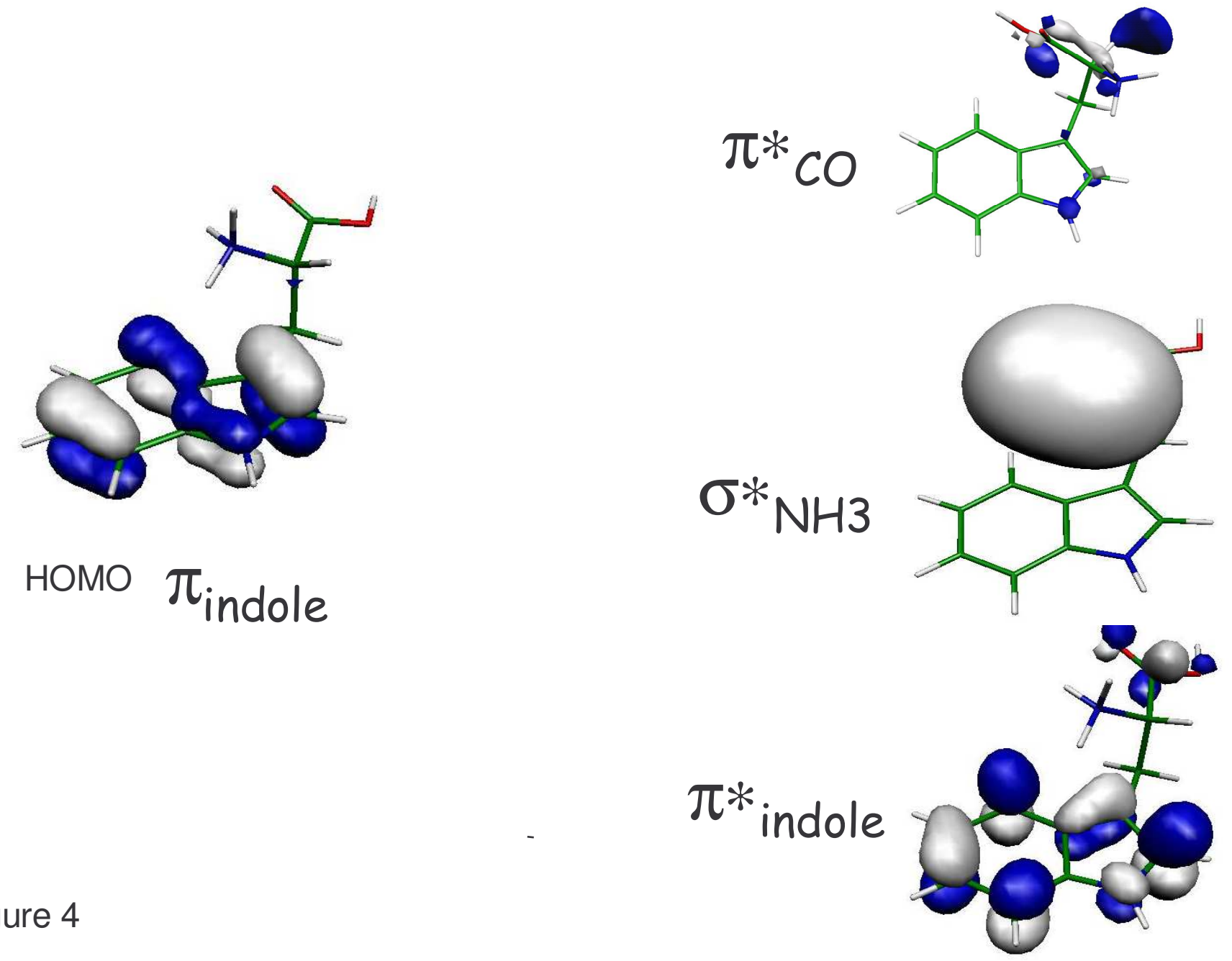

Figure 4

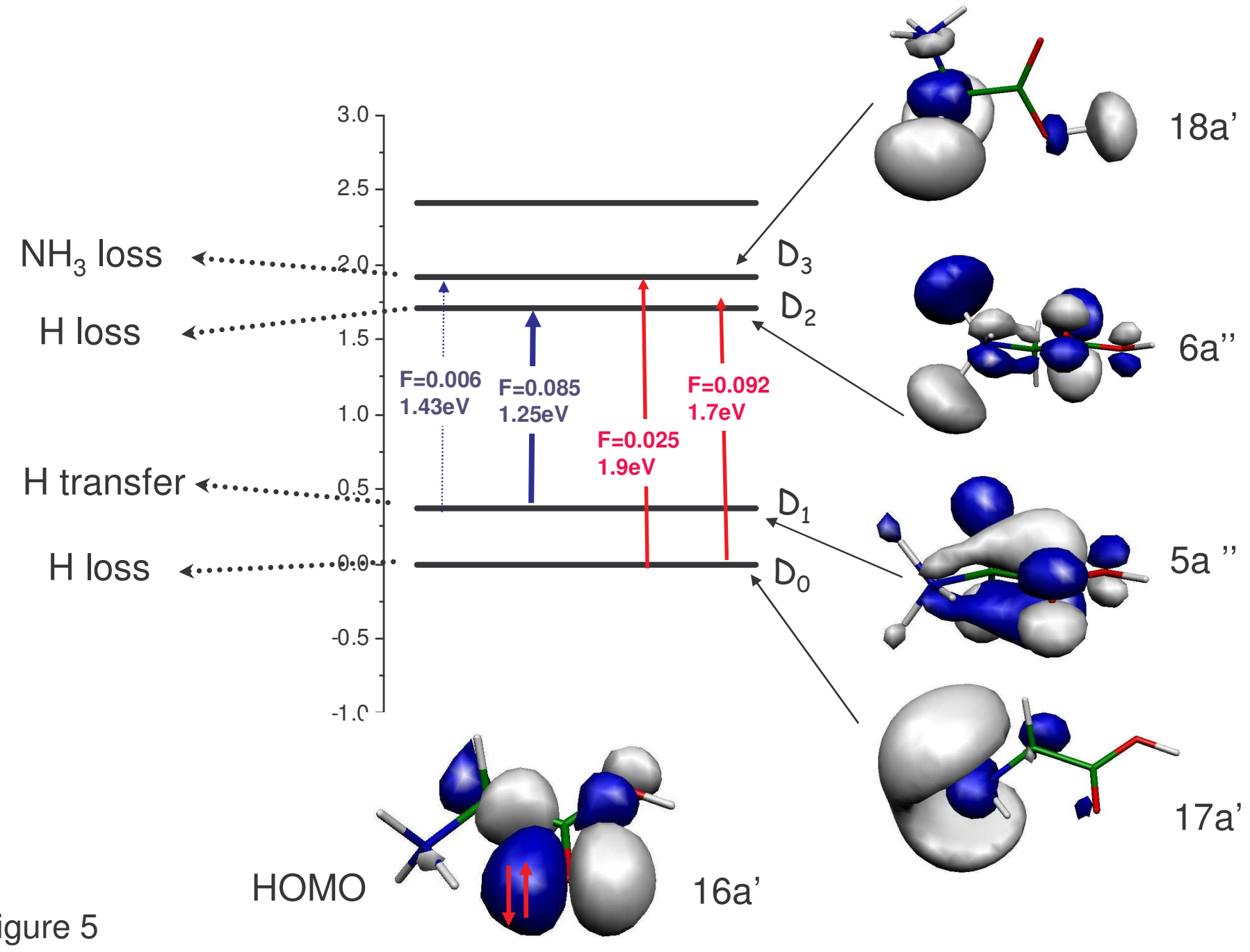

Figure 5

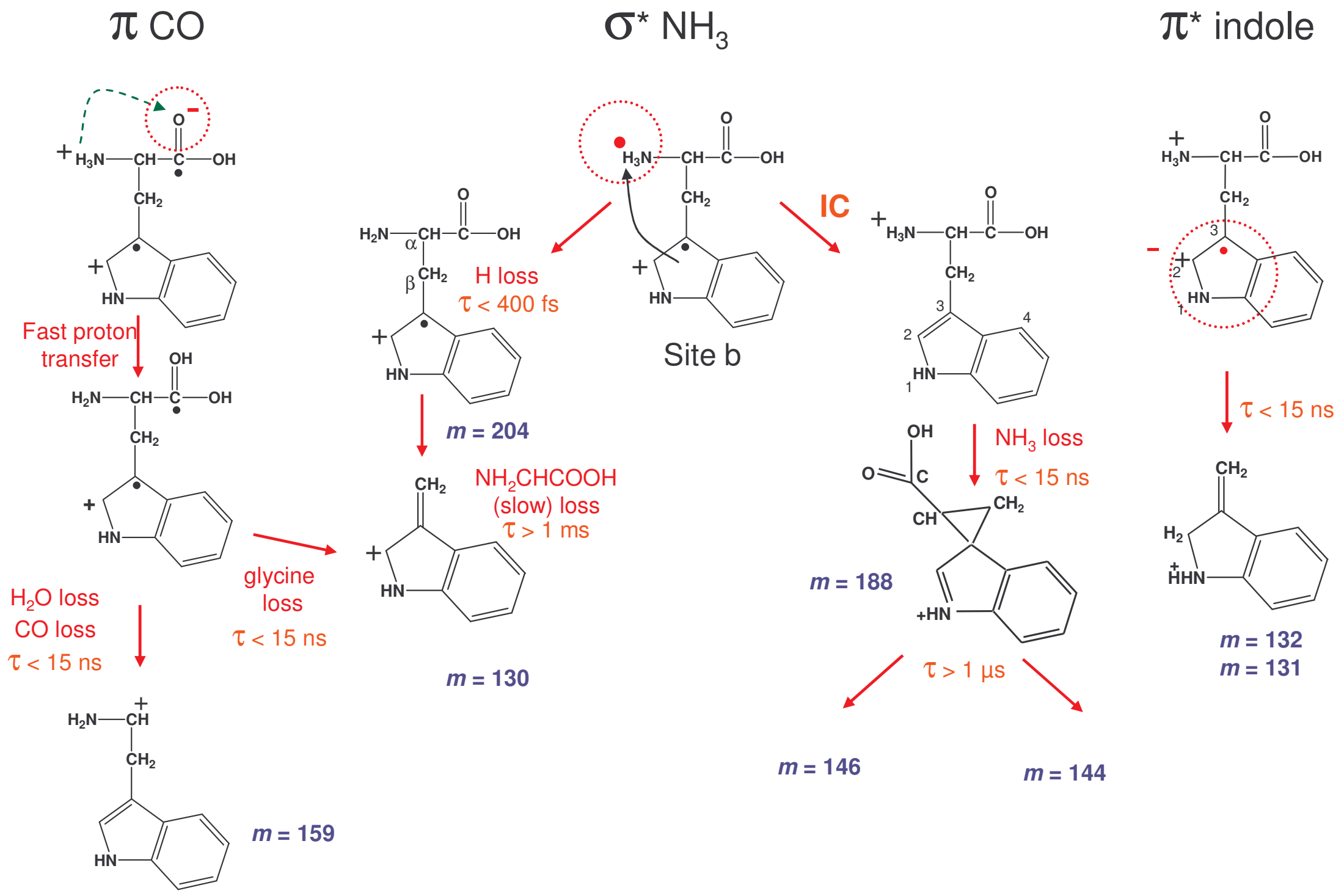

π CO
σ* NH3
π* indole
Fast proton transfer
H2O loss
CO loss
τ < 15 ns
glycine loss
τ < 15 ns
m = 159
H loss
τ < 400 fs
m = 204
NH2CHCOOH (slow) loss
τ > 1 ms
m = 130
Site b
IC
NH3 loss
τ < 15 ns
m = 188
τ > 1 μs
m = 146
m = 144
τ < 15 ns
m = 132
m = 131